\documentclass[12pt]{article}
\usepackage{epsf,epsfig,bbm}
\def\beq{\begin{equation}}
\def\eeq{\end{equation}}
\def\beqar{\begin{eqnarray}}
\def\eeqar{\end{eqnarray}}
\def\ga{\mathrel{\raise.3ex\hbox{$>$\kern-.75em\lower1ex\hbox{$\sim$}}}}
\def\la{\mathrel{\raise.3ex\hbox{$<$\kern-.75em\lower1ex\hbox{$\sim$}}}}

\begin{document}

\begin{titlepage}
\pagestyle{empty}
\baselineskip=21pt
\rightline{hep-ph/0608096}
\rightline{UMN--TH--2512/06, TPI--MINN--06/26}
\vskip 0.35in
\begin{center}
{\large{\bf The Fate of SUSY Flat Directions and their Role in Reheating
}}
\end{center}
\begin{center}
\vskip 0.05in
{{\bf Keith A.~Olive}$^{1}$
 and {\bf Marco Peloso}$^2$
\vskip 0.05in
{\it
$^1${William I Fine Theoretical Physics Institute, \\
University of Minnesota, Minneapolis, MN 55455, USA}\\
$^2${School of Physics and Astronomy, \\
University of Minnesota, Minneapolis, MN 55455, USA}\\
}}
\vskip 0.35in
{\bf Abstract}
\end{center}
\baselineskip=18pt \noindent

We consider the role of supersymmetric flat directions in reheating the Universe
after inflation. One or more flat directions can develop large vevs during inflation, 
which can potentially affect reheating by slowing down scattering 
processes among inflaton decay products or by coming to dominate 
the energy density of the Universe. Both effects occur only if flat 
directions are sufficiently long-lived. The computation of their perturbative 
decay rate, and a simple estimate of their nonperturbative decay  
have led to the conclusion that this is indeed the case. 
In contrast, we show that flat directions can decay quickly through nonperturbative 
channels in realistic models. The mass matrix for MSSM excitations around 
flat directions has nondiagonal entries, which vary with the phase of the (complex) 
flat directions. The quasi-periodic motion of the flat directions results in a strong 
parametric resonance, leading to the rapid depletion of the flat direction 
within its first few rotations. This may preclude any significant role for the flat 
directions in reheating the Universe after inflation
in models in which the inflaton decays perturbatively.

\vfill
\vskip 0.15in
\leftline{August 2006}
\end{titlepage}
\baselineskip=18pt

\section{Introduction}

One of the most attractive mechanisms for producing the baryon asymmetry of the Universe,
is the out-of-equilibrium decay of coherent scalar field oscillations 
along nearly $F$- and $D$-flat directions
of the scalar potential in supersymmetric theories \cite{ad,lin}.  
The minimal supersymmetric standard model contains many of these
flat directions \cite{tony} and one or 
more can be expected to be excited during inflation \cite{fluct,eeno}. 
There are many consequences of the evolution of  large vevs along flat directions \cite{em}.
If, in the context of a grand unified theory (GUT), there exists a
scalar operator $\mathcal{O}$, which violates baryon number and 
for which $\langle \mathcal{O} \rangle \ne 0$, the CP violating decay
of the flat direction will produce a large baryon asymmetry. 
Alternatively, any operator violating lepton number could be used to 
generate a net lepton asymmetry \cite{cdo}
which is subsequently converted to a baryon asymmetry by sphaleron processes.

Neglecting nonperturbative effects, one typically finds that the
scalar field oscillations of a flat direction persist after the decay of the inflaton
\cite{eeno,cgmo,am}. Indeed, quite generally, the Universe becomes dominated
by flat direction oscillations until these decay.  It has also been recently argued \cite{am}
that flat directions can delay the thermalization of the inflaton decay products
(resulting in a low reheat temperature) by providing a large mass to gauge bosons through their vev, and thus suppressing the rates of the scattering processes needed for thermalization.

Here, we examine the effects of preheating on the fate of flat directions. We find that nonperturbative effects can be important and lead to the rapid decay of the flat directions, long before they can dominate the expansion rate of the Universe. Similarly, the rapid decay may preclude any delay in the thermalization process after the decay of the inflaton. These effects are due to mixing between different MSSM excitations around many different sets of flat directions\footnote{Two or more mutually
non-exclusive flat directions are required. As we show below, only in exceptional cases does
a single flat direction exclude all others from being excited.},
and have been overlooked in the literature so far.

We will work in the context where the inflaton, $\psi$, couples to standard model
fields only through gravity\footnote{Reheating in the context of nonperturbative inflaton decay has been discussed in~\cite{rehnonpert}.}.  We will also assume for simplicity that
there is only a single mass scale associated with the inflaton potential set by the COBE 
normalization \cite{cobe},
with $m_\psi \sim H_I \sim 10^{-7} M_P$, where $H_I$ is the
Hubble parameter during inflation and $M_P$ is the Planck scale.  
We also assume that after inflation,
$\phi$ oscillates with a Planck scale amplitude.  
The energy density in oscillations can be written as
\beq
\rho_\psi = m_\psi^2 \psi^2 = m_\psi^2 M_P^2 (R_\psi/R)^3
\label{rhopsi}
\eeq
where $R_\psi$ is the Robertson-Walker scale factor when
oscillations begin.  For $R > R_\psi$, the Universe expands 
as a matter dominated Universe with Hubble parameter, 
$H \simeq m_\psi (R_\psi/R)^{3/2}$.
Inflatons decay at $R_{d\psi} \simeq (M_P/m_\psi)^{4/3} R_\psi$ when their
decay rate $\Gamma_\psi = m_\psi^3/M_P^2 \simeq
H$. After inflaton decay, the Universe is radiation dominated and $\rho_{r\psi}
\simeq m_\psi^{2/3} M_P^{10/3} (R_\psi/R)^4$, where $\rho_{r\psi}$ is the
energy density in the radiation produced by inflaton decay.

Thermalization occurs when the relevant scattering rates become comparable 
to the Hubble parameter.  Naive estimates indicate that this may be delayed
due to suppressed cross sections of the form $\sigma \sim (\alpha^2/m_\psi^2)(R/R_{d\psi})^2$
where $\alpha$ is a characteristic coupling strength (presumably of order the
gauge fine structure constant) and the suppression is due to the initially large 
particle energies produced in the decay ($E \sim m_\psi$) which redshift to 
lower energies as the Universe expands.  The resulting reheat temperature in
this case would be expected to be low, $T_R \sim 10^4$ GeV \cite{eeno}.
However it was shown \cite{ds} that forward scattering in 2 $\to$ 3 interactions, though
of higher order in $\alpha$, lead to rapid thermalization.  It was estimated that
the inelastic rates were roughly $\Gamma_{inel} \sim \alpha (M_P/m_\psi)n_\psi \sigma$  
in which case thermalization occurs almost immediately
after decay and
restores the simple estimate $T_R \sim (\Gamma_\psi M_P) \sim 10^8$ GeV.
This will be true so long as $\alpha^3 > m_\psi/M_P$.

In supersymmetric theories, quantum fluctuations can drive 
several scalar fields to large expectation values along flat directions.  
In fact, in supergravity models with a Heisenberg symmetry \cite{BG}, such as no-scale
supergravity \cite{ns}, we expect to generate vev's of order $\phi_0 \sim M_P$
for any flat direction  not involving stops \cite{gmo}.  This vev may be smaller
if the flat direction is regulated by non-renormalizable operators \cite{cgmo}.

When the Hubble parameter after inflation is of order the supersymmetry 
breaking scale $m_\phi$, sfermion oscillations along the flat direction will begin.
This occurs at $R = R_\phi \simeq (m_\psi/m_\phi)^{2/3} R_\psi$.
The energy density in the oscillations of the flat direction field characterized by $\phi$ is
given by
\beq
\rho_\phi = m_\phi^2 \phi^2 = m_\phi^2 \phi_0^2 (R_\phi/R)^3 \simeq m_\psi^2 \phi_0^2 (R_\psi/R)^3
\eeq
Typically, oscillations of $\phi$ will begin before inflaton decay, since $(R_{d\psi}/R_\phi)
\sim M_P^{4/3} m_\phi^{2/3}/m_\psi^2 \sim 10^3$.

We note that although most standard model (SM) fields are massive
when the flat direction is excited, inflaton decay is unimpeded if the inflaton can decay
directly into the light fields making up the flat direction.  
Nevertheless, the scattering rates will be
suppressed due to the large masses of exchanged particles \cite{am}.
To compute the thermalization rate of the decay products,
we once again consider the inelastic 2 $\to$ 3 scatterings
by replacing $\sigma$ with $(\alpha^2/\phi^2)$ and $\phi^2 = (\phi_0^2 m_\psi^2/m_\phi^2)
(R_\psi/R)^3$. In this case,  thermalization is not suppressed unless
$\alpha <m_\psi^{5/3}/m_\phi^{2/3} M_P$.  For $\alpha^2 \sim 10^{-3}$,  one finds again
a reheat temperature of order $T_R \sim \alpha^{3/2} m_\phi M_P/m_\psi \sim 10^8$ GeV.
Note that we have assumed $\phi_0 = M_P$.  For lower $\phi_0$, reheating at this
stage will always be given by the instantaneous rate yielding $T_R \sim 10^8$ GeV.
However as we will see, the energy density of the Universe will 
become dominated by the oscillations along the flat direction
and final thermalization will depend on the decay of the flat direction vev\footnote{As we show below,
nonperturbative effects associated with mixing dramatically alter this conclusion.}.

If inflaton decay occurs before the decay of flat direction,  
the radiation produced by inflaton decay will redshift 
faster than the flat direction oscillations.  It is of interest, therefore,
to compare the energy density in inflaton decay products
with that in the flat direction oscillations at the time of decay (at $\phi = \phi_d$).
The ratio of energy density in radiation to oscillations is
\beq
\frac{\rho_{r\psi}}{\rho_\phi} = \left( \frac{M_P^{10/3}}{m_\psi^{4/3} \phi_0^2} \right) \left( \frac{R_\psi}{R} \right)
\eeq
 The oscillations end when the decay rate, $\Gamma_\phi \sim m_\phi^3/\phi^2 \sim H$,
 from which we deduce that $R_{d\phi} = (m_\psi^{7/15} \phi_0^{2/5} M_P^{2/15} /m_\phi) R_\psi$.
 At this value of $R$, we see that inflaton decays dominate
 the energy density only if $\phi_0 < m_\phi^{5/12} M_P^{4/3} / m_\psi^{3/4} \sim 10^{-2} M_P$.
 For larger values of the initial vev along the flat direction, 
 reheating and thermalization is entirely determined by the dynamics of the flat direction \cite{cgmo}.
 In this case, we expect a baryon asymmetry of order $\mathcal{O}(1)$ \cite{ad,lin}
 unless it is diluted by other processes such as moduli decay or the vev is regulated by
 non-renormalizable or GUT scale operators \cite{cgmo}.
 
 For relatively large $\phi_0$, the oscillations of the flat direction
 come to dominate the energy density.  In this case the decay of $\phi$ is
 determined by $\Gamma_\phi = (m_\phi^5 / \phi_0^2 m_\psi^2) (R / R_\psi)^3 = 
 (m_\psi \phi_0 / M_P) (R_\psi/R)^{3/2} = H$. The scale factor at decay is
 $R_{d\phi} = (\phi_0^{2/3} m_\psi^{2/3} / m_\phi^{10/9} M_P^{2/9} ) R_\psi$.
 Because this decay occurs relatively late,  the reheat temperature is
 suppressed, $T_R \sim m_\phi^{5/6} M_P^{1/6}$  $\sim 10^6$~GeV.

The above argument for the decay of $\phi$ ignores the possibility that the flat directions
decay on a much quicker timescale in a nonperturbative fashion (preheating). In the following,
we show that these effects can be indeed dominant, and lead to the quick depletion of the flat directions
within their first ($\sim 5-10$) rotations. If so, the Universe remains dominated by inflaton oscillations until inflaton decay. Reheating in this case is produced by inflaton decays, and the baryon asymmetry is expected to be large unless $\phi_0$ is regulated by non-renormalizable operators or dilution due to moduli decay occurs.

The plan of the paper is as follows. In Section \ref{sec2}, we discuss an argument, based on a toy model of preheating, which has been used to claim that preheating along flat directions
is negligible. In the remainder of that Section, and in Section \ref{sec3}, we explain why 
this conclusion might not remain valid when realistic models are considered. In Section \ref{sec4}, we present an      explicit computation of the actual nonperturbative decay. Section \ref{sec5} contains our conclusions.

\section{Nonperturbative decay of the flat direction} \label{sec2}

In this section, we will first consider the case of two complex fields with a global 
U(1) symmetry.  We will see later that the time evolution of the 
mass matrix along the flat direction leads to preheating.
However, when the U(1) is gauged, the fields involved in the preheating are simply the massive Higgs and its related Goldstone boson eaten by the gauge field.  We next consider a case
with two flat directions involving four complex fields.  In this case, we find that in addition
to the four real degrees of freedom associated with the decoupled flat directions, and the 
massive Higgs and Goldstones fields, we are left with two real degrees of freedom for which
preheating is found to occur.

We saw in the previous section that when preheating is neglected,
the flat direction comes to dominate over the
inflaton decay products, if its initial amplitude is sufficiently large, $\phi_0 \ga 10^{-2} M_p \,$.
In this case, the thermal bath which is relevant for the following history of the Universe is the one
formed at the decay of the flat direction. Most existing studies focus only on the perturbative
(single quanta) decay of $\phi \,$. The possibility of a nonperturbative decay along flat directions was considered in~\cite{asc,poma} where an important difference between this case and nonperturbative inflaton decay at preheating was pointed out. In the latter case, the inflaton is typically assumed to be a scalar field oscillating about the minimum of its potential, where its amplitude vanishes. Flat directions are instead complex fields. Nonrenormalizable operators (as well as susy breaking masses), give rise to terms in the potential which also depend on the the phase $\sigma$ of the flat direction ($\phi = \vert \phi \vert \, {\rm exp} \left( i \sigma \right) \,$). This in general leads to a time evolution of $\sigma \,$, which can be seen as out of phase oscillations of the real an imaginary parts of the flat direction; hence, $\phi$ never vanishes during these oscillations. 

The general picture was originally discussed in~\cite{ad}, where the ``angular'' motion of $\phi$ is associated to a baryon number charge. A nonrenormalizable operator is responsible for an initial $\dot{\sigma} \neq 0 \,$. Subsequently, the field evolves to a lower amplitude, where the dominant potential term $m_\phi^2 \vert \phi \vert^2$ is $\sigma$ independent. This results in an approximately elliptical motion of the flat direction in its complex plane (the amplitude slowly decreases due to the expansion of the Universe, so that $\phi$ is actually spiraling down towards the origin). For the following estimates, we can assume that the amplitude of $\phi$ oscillates between the maximum $\phi_0$ and the minimal value $\epsilon \phi_0 \,$ over a timescale $m_\phi^{-1} \,$. The value $\epsilon = 0$ corresponds to  straight line radial motion of $\phi$ in the complex plane ($\dot{\sigma} = 0$); in this case the phase $\sigma$ can be rotated away, and $\phi$ oscillates as a real field. In the supersymmetric case, the value of $\epsilon$ depends on the actual potential, and on the initial conditions for $\phi$. Values $10^{-3} \la \epsilon \la 10^{-1}$ are typically found \cite{poma}. 

In analogy to what has been computed for inflationary preheating, one can consider~\cite{asc,poma} the excitation of a single (complex) field $\chi$, due to its coupling with the background flat direction, $\phi$,
\begin{equation}
\Delta V = g^2 \vert \phi \vert^2 \vert \chi \vert^2
\label{pot1}
\end{equation}
This interaction leads to a time dependent effective mass for the real and imaginary components of $\chi$,
\begin{equation}
m_{{\rm eff}, \chi}^2 = m_\chi^2 + g^2 \vert \phi \left( t \right) \vert^2
\label{mass1}
\end{equation}
(the ``bare'' mass $m_\chi$ is expected to be of order the electroweak scale, and can be neglected for these considerations).
The nonperturbative decay of the flat direction takes place whenever the frequency of the quanta of $\chi$ varies nonadiabatically, $\dot{\omega} \ga \omega^2 \,$. For relatively small momenta $p$, one finds $\omega^2 = p^2 + m_{{\rm eff}, \chi}^2 \simeq g^2 \vert \phi \left( t \right) \vert^2 \,$. The ratio $\dot{\omega} / \omega^2$ is maximized when the amplitude of $\phi$ is minimal. When this occurs, one finds $\dot{\omega} / \omega^2 \simeq g \epsilon \phi_0 m_\phi / g^2 \epsilon^2 \phi_0^2 \,$. Therefore, $\phi$ decays nonperturbatively only if
\begin{equation}
\epsilon \la \frac{m_\phi}{g \phi_0}
\end{equation}
Since, $m_\phi$ is of order the electroweak scale, while $\phi_0$ is close to the Planck scale, even a very small deviation from radial motion invalidates this inequality. Based on this consideration~\cite{asc,poma}, it was concluded~\cite{am} that the flat direction decays only perturbatively, leading to 
a suppressed reheating temperature and a solution to the gravitino problem.

However, this conclusion strongly depends on the coupling~(\ref{pot1}) assumed in these analyses, and in the resulting mass term~(\ref{mass1}). In concrete cases, the coupling is typically more complicated. 
For instance, the interaction may couple the real and imaginary components of $\chi \,$, or could couple $\chi$ to some other field. Furthermore, the coefficients of these couplings may depend also on the phase $\sigma$ of the flat direction, in addition to its amplitude. As we argue in the next Section, this is actually what happens for realistic MSSM flat directions. For definiteness, 
consider a flat direction involving two complex scalar fields with a global U(1) symmetry and a 
potential
\beq
V = \frac{g^2}{8} \left( |\Phi_1|^2 - |\Phi_2|^2 \right)^2
\label{toyv}
\eeq
which has the same structure as the MSSM potential from the $D-$ terms which we will discuss below.
If we expand the fields about their background value, $\phi$, we can write
\beq
\Phi_1 = \phi + (\xi + \chi)
\eeq
and
\beq
\Phi_2 = \phi + (\xi - \chi)
\eeq
which when inserted into (\ref{toyv}) becomes
\begin{equation}
V = \Delta V = \frac{g^2}{8} \left(4 \, {\rm Re} (\phi \, \chi^*) + 4 \, {\rm Re} (\xi^* \, \chi \right)  )^2
\label{pot2}
\end{equation}
This interaction couples the real and the imaginary parts of $\chi$ (which we denote by $\chi = \left( \chi_1 + i \,\chi_2 \right) / \sqrt{2} \,$), and, consequently, the effective mass~(\ref{mass1}) is now replaced by a $2 \times 2$ nondiagonal mass term
\begin{equation}
\Delta V = \left( \chi_1 \,,\, \chi_2 \right)_i \, {\cal M}_{ij}^2 \left( \chi_1 \,,\, \chi_2 \right)_j \;\;\;,\;\;\; 
{\cal M}^2 \equiv 
\left( \begin{array}{cc}
m_{\chi_1}^2 + f^2 \cos^2 \sigma & f^2 \cos \sigma \sin \sigma \\
f^2 \cos \sigma \sin \sigma & m_{\chi_2}^2 +f^2 \sin^2 \sigma
\end{array} \right)
\label{mass2}
\end{equation}
where $f \equiv \sqrt{2} g \vert \phi \vert$. Neglecting the small electroweak masses $m_{\chi_1} ,\, m_{\chi_2} \,$, one linear combination of $\chi_1 ,\, \chi_2$ is massless, while the other one has mass $f^2 \,$. Since this eigenmass only depends on the amplitude of $\phi \,$, one may be tempted to conclude that the above considerations still apply, and that nonperturbative effects are negligible in this case as well.

However, the nature of the massive and massless combinations is time dependent; when $\sigma = 0$ the massive combination coincides with $\chi_1 \,$, while $\chi_2$ is massless. The real and imaginary components interchange their role every quarter of rotation. We will see that this time variation is also a source of nonperturbative particle production, which, being proportional to  $\dot{\sigma}$, is not suppressed as $\epsilon$ increases. In particular, it takes place even if the eigenmasses are constant or vary only adiabatically. As we show in Section~\ref{toy}, strong nonperturbative production takes place even if the motion of $\phi$ is perfectly circular, and $\vert \phi \vert$ is constant. 

Now if we gauge the U(1),  as the forthcoming analogies with the MSSM will require,
we can show that the massive combination of $\chi_1 ,\, \chi_2$ is in fact a Higgs boson
while the other combination is the Goldstone boson.  To see this, consider the coupling between
the gauge boson and the scalar field excitations emerging from the covariant derivatives of $\Phi_i$. This coupling could be removed by an infinitesimal $U \left(1 \right)$ gauge transformation, with the gauge parameter $\lambda$ satisfying
\begin{eqnarray}
\partial_{\vec x} \lambda &=& \partial_{\vec x} \left[ \frac{\sin \sigma \, \chi_1 - \cos \sigma \, \chi_2}{\sqrt{2} \vert \phi \vert} \right] \nonumber\\
\partial_0 \lambda &=& \partial_0 \left[ \frac{\sin \sigma \, \chi_1 - \cos \sigma \, \chi_2}{\sqrt{2} \vert \phi \vert} \right] - \frac{\sqrt{2} \dot{\sigma}}{\vert \phi \vert} \left( \cos \sigma \, \chi_1 + \sin \sigma \, \chi_2 \right)
\label{lambdasys}
\end{eqnarray}
In the standard higgs mechanism, $\dot{\sigma} = 0$, and one can completely decouple the gauge boson 
from the neutral excitations by choosing
\begin{equation}
\lambda = \frac{\sin \sigma \, \chi_1 - \cos \sigma \, \chi_2}{\sqrt{2} \vert \phi \vert}
\label{lambdasol}
\end{equation}
The unitary gauge eliminates the massless Goldstone boson and the remaining massive higgs excitation is decoupled. In the present case, (\ref{lambdasol}) is clearly not a solution of~(\ref{lambdasys}); however, it is still convenient to make this gauge choice, since it eliminates the massless neutral excitation, leading to the following quadractic action for the fluctuations
\begin{equation}
\vert D_\mu \Phi_i \vert^2  - V \supset \frac{1}{2} \left( \partial_\mu \chi_m \right)^2 + \frac{1}{2} \left( \partial_\mu \sigma \right)^2 \chi_m^2 + 2 g^2 \vert \phi \vert^2 A_\mu^2 - 4 g \vert \phi \vert \dot{\sigma} A_0 \chi_m - 2 g^2 \vert \phi \vert^2 \chi_m^2
\label{gauge}
\end{equation}
where $\chi_m = \left( \cos \sigma \chi_1 + \sin \sigma \chi_2 \right)$ is the massive higgs, while the orthogonal massless scalar has been absorbed in the longitudinal part of $A_\mu$.  The mass matrix 
for the $\left\{ \phi_m ,\, A_0 \right\}$ system is still nondiagonal, due to the time evolution of $\sigma \,$.
However, the nature of the corresponding eigenstates does not rotate in field space, and preheating effects can be neglected as long as $\dot \sigma \ll \vert \phi \vert \,$.

Even in the gauged case, significant preheating from the rotation of the mass matrix will take place
in presence of additional coupled degrees of freedom beyond the Higgs, Goldstone pair. This will happen in presence of $2$ or more flat directions. To see this, consider a second toy model model with 2 flat directions characterized by four complex fields, $\Phi_i$
with a potential  
\beq
V = \frac{g^2}{8} \left( q \, |\Phi_1|^2 - q \,  |\Phi_2|^2 + q' \, |\Phi_3|^2 - q' \, |\Phi_4|^2 \right )^2
\eeq
where we have assigned the fields $\Phi_{1,2}$ and $\Phi_{3,4}$ charges
$\pm q$ and $\pm q'$, respectively. We expand $\Phi_{1,2}$ as above, and
\begin{equation}
\Phi_3 = \phi' + \left( \xi' + \chi' \right) \;\;\;,\;\;\; \Phi_4 = \phi' + \left( \xi' - \chi' \right) \nonumber\\
\end{equation}
and, for simplicity, we take $q' \vert \phi' \vert = q \vert \phi \vert \,$. In this case, the part of the potential quadratic in the fluctuations reads
\begin{equation}
\Delta V = \frac{f^2}{2} \left( \cos \sigma \, \chi_1 + \cos \sigma' \, \chi_2' + \sin \sigma \, \chi_1
+ \sin \sigma' \, \chi_2' \right)^2
\label{dvtoy2}
\end{equation}
(where $\sigma'$ is the phase of $\phi'$).

The corresponding mass matrix has one massive and three massless eigenstates, whose nature
changes with time in the $\left\{ \chi_1 ,\, \chi_2 ,\, \chi_1' ,\, \chi_2' \right\}$ basis, as the two flat directions rotate. We have verified explicitly that all four states are produced at preheating (the computation is analogous to the one for the mass matrix~(\ref{mass2}), which is described in Section 4), as long as $\dot{\sigma} \neq \pm \dot{\sigma}^{\prime} \,$. When the $U \left( 1 \right)$ symmetry is gauged, one massless combination (the Goldstone boson) is eliminated; however, 
the three remaining eigenstates still rotate in field space, and the corresponding eigenvalues are produced during preheating. In the special case when $\dot{\sigma}^\prime = \dot{\sigma}$ ($\dot{\sigma'} = - \dot{\sigma}$), we are effectively back to a single flat direction, and indeed we note that only the two combinations $\chi_1 + \chi_1'$ and $\chi_2 + \chi_2'$ ( $\chi_2 - \chi_2'$) are entering in $\Delta V$; in this case, the result (or lack thereof) for preheating is identical to the one for the toy model~(\ref{toyv}).

We conclude this Section with a note on the nonlinear interactions of the scalar modes and on their possible effects on preheating. The excitation of fields produced at preheating is typically encoded in a large variance for these fields. However, if for example a self-interaction $\lambda \chi^4$ is present in the potential, then energy considerations preclude the formation of a large variance. Therefore, we would conclude that such a field $\chi$ cannot be significantly excited during preheating. This is particularly true if $\lambda$ is as large as typical gauge couplings~\cite{bdps}. These considerations, however, do not affect the toy models just discussed. From eq.~(\ref{pot2}) we see that the nonlinear interactions of $\chi$ involve the field $\xi \,$, which is not produced at preheating. Energy considerations then only limit the growth of the variance of $\xi$ (which could be produced when rescattering effects of $\chi$ are taken into account), but $\langle \chi^2 \rangle$ itself can have
significant growth, as long as $\xi$ remains sufficiently small. An analogous situation takes place in the second toy model, where the nonlinear interactions involve the fields $\xi$ and $\xi'$ which are not produced at preheating.

\section{The mass matrix for fluctuations about a flat direction} \label{sec3}

The mass matrix ${\cal M}^2$ introduced in the previous Section (namely the one in eq.~(\ref{mass2}) without the small susy breaking masses $m_{\chi i}$) is precisely the one found for fluctuations around a MSSM flat direction, when one considers the part of the potential coming from $D-$terms:
\begin{equation}
V_D = \sum_a \frac{g_a^2}{2} D^a D^a \;\;\;,\;\;\;
D^a = \sum_i X_i^* T^a X_i
\label{dterm}
\end{equation}
where the index $a$ runs over the generators of the gauge group, and $X_i$ are the MSSM fields charged under that group element. To find the mass term, consider the  MSSM excitations around a generic flat direction with vev $\phi \,$. In this way, each MSSM field $X_i$ is either of the form $\phi + \delta X_i$ (if the field takes part in the flat direction) or $\delta X_i$. By definition, each $D^a$ vanishes when evaluated on a flat direction, that is when the fluctuations $\delta X_i$ are set to zero. For this reason, $D^a$ is at least linear in the excitations,
\begin{eqnarray}
D^a &=& \phi^* \sum_i c_i \delta X_i + {\rm h. c.} + {\rm O} \left( \delta X^2 \right) \nonumber\\
&=& 2 \vert \phi \vert \left[ \cos \sigma \, {\rm Re} \left( \sum_i c_i \delta X_i \right) +
\sin \sigma \, {\rm Im} \left( \sum_i c_i \delta X_i \right) \right] + {\rm O} \left( \delta X^2 \right)
\label{dalinear}
\end{eqnarray}
where $c_i$ are numerical coefficients dependent on the flat direction and on the generator considered\footnote{The linear term~(\ref{dalinear}) arises from mixed terms in~(\ref{dterm}), in which we take the background $\phi$ from one of the two $X_i$ and the excitation from the other.}. It is clear that squaring~(\ref{dalinear}) results in general in a mass matrix which depends both on the amplitude and the phase of $\phi$, and which is non-diagonal in field space. Moreover, if we go to a basis constructed from the linear combinations entering in the linear term, the resulting mass matrix will be block diagonal, with all nonvanishing blocks proportional to ${\cal M}^2$.

\subsection{A single flat direction}

It is instructive to see this explicitly for  a specific example. We consider the $LLE^c$ flat direction, and for definitness, we give a vev to
\begin{equation}
\langle \nu_e \rangle = \langle \mu \rangle = \langle \tau^c \rangle = \phi
\end{equation}
In addition, we define $\delta X = \left( \delta X_R + i \delta X_I \right) / \sqrt{2}$ to be the fluctuation of each MSSM field about this flat direction. The mass matrix from the $D-$terms involves $10$ coupled real fields\footnote{We are focusing here only on the part of the potential coming from $D-$terms. The part of the potential from the $F-$ terms gives a mass matrix for a separate set of MSSM excitations, and it is diagonal in field space and $\sigma-$independent.}. They are the real and imaginary parts of the fluctuations of the three fields involved in the flat direction (system $I$), plus the real and imaginary parts of $\delta e$ and $\delta \nu_\mu \,$ (system $II$). The two systems are actually decoupled at the quadratic level, since the mass matrix for the fields in the first system comes from the diagonal $D^3$ (isospin) and $D'$ (hypercharge) $D-$terms, while the one for the second system arises from the SU$(2)$ off diagonal generators.

The off diagonal structure for the first system arises in the 
\begin{equation}
\left\{ \frac{\left( \nu_e - \mu \right)_R}{\sqrt{2}} ,\, 
\frac{\left( \nu_e - \mu \right)_I}{\sqrt{2}} ,\,
\frac{\left( \nu_e + \mu - 2 \tau^c \right)_R}{\sqrt{2}} ,\,
\frac{\left( \nu_e + \mu - 2 \tau^c \right)_I}{\sqrt{2}} ,\,
\frac{\left( \nu_e + \mu + \tau^c \right)_R}{\sqrt{2}} ,\,
\frac{\left( \nu_e + \mu + \tau^c \right)_I}{\sqrt{2}} 
\right\}
\label{base1}
\end{equation}
basis, where the mass matrix reads
\begin{eqnarray}
M_I^2 = \left( \begin{array}{ccc}
g^2 \vert \phi \vert^2 {\cal M}^2 & 0 & 0 \\
0 & 3 g^{' 2} \vert \phi \vert^2 {\cal M}^2 & 0 \\
0 & 0 & 0
\end{array} \right)
\end{eqnarray}
We note that each entry in this mass matrix is actually a $2 \times 2$ matrix. The first nonvanishing block is the mass matrix for the coupled system composed by a heavy higgs and the Goldstone boson
which provides the longitudinal component of $W_3$. The second block is instead the mass matrix mixing another heavy higgs with the goldstone boson which provides the longitudinal component to $B$. The remaining two massless eigenstates are the actual real and imaginary components of the flat directions, and are decoupled at the quadratic level.

For the second set of fields, in the basis,
\begin{equation}
\left\{ \frac{\left( e + \nu_\mu \right)_R}{\sqrt{2}} ,\,
\frac{\left( e + \nu_\mu \right)_I}{\sqrt{2}} ,\,
\frac{\left( e - \nu_\mu \right)_I}{\sqrt{2}} ,\,
\frac{\left( - e + \nu_\mu \right)_R}{\sqrt{2}} 
\right\}
\label{base2}
\end{equation}
we find instead
\begin{eqnarray}
M_I^2 = \left( \begin{array}{cc}
g^2 \vert \phi \vert^2 {\cal M}^2 & 0 \\
0 & g^2 \vert \phi \vert^2 {\cal M}^2
\end{array} \right)
\label{sysii}
\end{eqnarray}
where each block is the mass matrix between a heavy charged field and the Goldstone boson providing the longitudinal component to $W_1$ and $W_2 \,$, respectively.

In summary, the $LLE^c$ flat direction breaks SU$(2) \times$ U$(1)$ completely. For each broken symmetry, there is a massive (higgs) and a massless (Goldstone) excitation; in addition, the are two massless excitations (whose nature does not change with time) which represents the actual fields associated with the flat direction. This pattern ($1$ massive and $1$ massive field for each broken symmetry $+$ 2 additional massless fields) applies for each flat direction.

This shows that preheating from mixing does not occur if we have a single flat direction. The field redefinitions~(\ref{base1}) and (\ref{base2}) are constant in time, and do not lead to particle production.
After these redefinitions, we are left with a sum of systems analogous to the first  toy model
discussed in the previous Section, where the quick rotation in field space is eliminated once each
goldstone is absorbed in the longitudinal component of the corresponding gauge boson (cf. eq~(\ref{gauge}).) However, the quick rotation of the mass matrix in field space, and the corresponding preheating effect, takes place if two or more flat directions are excited, as we have discussed at the end of the previous Section (cf. the discussion after eq.~(\ref{dvtoy2})).

\subsection{Two or more flat directions.}

The flat direction considered above allows several other independent flat directions
to be excited simultaneously.  Note that this is not a general feature of all flat
directions.  For example, the $H_uH_d$ flat direction makes it very difficult to excite
other directions which remain F-flat.  However, there are many mutually non-exclusive directions,
including the one discussed in the previous subsection.

For example, let us consider the simultaneous presence of the two
flat directions $LLE^c$ and $QLD^c$. For definitness, we give vevs to
\begin{equation}
 \langle \mu \rangle = |\phi| ~~~ \langle \tau^c \rangle = |\phi|e^{i\theta} \qquad  
 \langle d_1 \rangle = |\phi^\prime| ~~~  \langle s_1^c \rangle = |\phi^\prime| e^{i\sigma} \qquad
  \langle \nu_e \rangle = \sqrt{|\phi|^2 + |\phi^\prime|^2}
\end{equation}
One can check explicitely that the scalar potential is both F- and D- flat along
this direction.  
Of the 5 potential phases, only the combinations ($\arg{\nu_e} + \arg{\mu} + \arg{\tau^c}$)
and ($\arg{\nu_e} + \arg{d_1} + \arg{s_1^c}$) are gauge invariant. 
Thus we are free to make the above assignments which will prove useful below. 
This combination of vevs will break the standard model
gauge group $SU(3)_c \times SU(2)_L \times U(1)_Y \to SU(2)_c$, ie. leaving
a color $SU(2)$ subgroup unbroken.  As a result, we expect 9 massive states
corresponding to Higgses, and 9 Goldstone modes.  
Our system is clearly more complicated now.  Nevertheless, in analogy with our
previous discussion, we can separate the excitations  which participate in the mass
matrix as follows.  The real and imaginary parts of the pairs $(d_2,s^c_2)$ and $(d_2,s^c_2)$
will split into two $4 \times 4$ blocks as in eqs. (\ref{base2}) and (\ref{sysii}), each containing
two massive Higgses and two Goldstones responsible for the longitudinal components of the 
four off-diagonal color generators ($T_{1,2}$ and $T_{4,5}$).  

Next, we can further isolate a $6 \times 6$ matrix for the fields, ${e}_R,  {e}_I, 
{\nu_\mu}_R,  {\nu_mu}_I, {u_1}_R,  {u_1}_I$.This matrix contains 2 Higgses which 
break the off diagonal generators of $SU(2)_L$ and their corresponding Goldstone modes.
Due to our choice of phase assignments, the entire matrix is phase independent and the
two remaining massless fields do not contribute to preheating.  

The remaining 10 fields, are the fluctuations of the fields with nonvanishing vevs, namely
${\nu_e}_R,  {\nu_e}_I, {\mu}_R,  {\mu}_I, 
{\tau^c}_R,  {\tau^c}_I, {d_1}_R,  {d_1}_I, 
{s_1^c}_R,  {s_1^c}_I$. Their mass matrix arises from the $D-$ terms corresponding to the broken gauge symmetries with diagonal generators (namely, hypercharge, isospin, and the sum of $T_3$ and $T_8$ color generators). For this reason, they form a system which is analogous to system $I$ described in the previous Subsection. Specifically, the three states  ${\nu_e}_i, \mu_i$, and ${d_1}_i$ do not enter in the mass matrix, due to the above phase assignment for the flat directions. The mass matrix for the remaining $7$ fields depends on the phases $\sigma$ and $\theta$. The overall system of $10$ states is characterized by $3$ massive eigenstates, $3$ Goldstones (which provide the longitudinal components to the gauge fields of the broken diagonal symmetries), and $4$ light physical states. 
Clearly, all the eigenstates are linear combinations of the $10$ states of the system. What is relevant for the present discussion, is that each of the $4$ light physical states has some coefficients in this linear combination which are proportional to the cosine and sine of the two phases $\sigma$ and $\theta \,$.
Therefore, these states are produced by preheating from mixing, in an analogous way to what happens in the two flat direction toy model of the previous Section.

It is interesting to note that the number of fields excited may be larger still.  For example,
a direction involving $U^cD^cD^c$ with  $\langle c^c_2 \rangle = \langle d_3^c \rangle 
=\langle b_1^c \rangle = \phi^{\prime\prime}$ is compatible with the previous
two flat directions discussed.  While the remaining $SU(2)_c$ subgroup is now broken,
The new direction involves nine new complex fields.  Of these 18 new degrees of freedom,
three are Higgses, three are Goldstones, two are decoupled, but 10 are light and are involved in 
mixing. 

We note that mutually non-exclusive flat directions are not restricted to simple
monomials involving three chiral superfields.  The direction $LLD^cD^cD^c$ is 
compatible with $QQQL$.  For example, giving vevs to
$\nu_e, \mu, d_1^c, s_2^c, b_3^c, u_2, d_3, c_1$, and $\tau$ would
involve 54 fields in the mass matrix.  With 12 Higgs and Goldstone modes, 
we are left with 30 fields which may participate in preheating!

We also wish to stress that these flat directions are expected to be excited
by scalar field fluctuations during inflation.  In that case, all mutually non-exclusive
directions will be excited.  In other words, if the $LLE^c$ direction we discussed in the previous subsection is excited, there is no way to prevent the other compatible flat directions
from acquiring large vevs as well.  

As we did at the end of Section~\ref{sec2}, we conclude also this Section with a note on nonlinear interactions. Contrary to what happens in the toy model with two flat directions, we expect that, for a generic flat direction, the fields produced at preheating have large (namely, of gauge strength) self-interactions, as well as large interactions with some of the other MSSM fields. It is well known from models of inflaton preheating that large self-interactions can strongly limit the excitation of a field at preheating. For instance, if we couple a field $\chi$ to the inflaton $\psi$ through a quartic $g^2 \psi^2 \chi^2$ interaction, strong preheating is expected through parametric resonance. However, the production of $\chi$ is strongly suppressed in the presence of a quartic interaction $\lambda \chi^4$, with large coupling $\lambda$. Unfortunately, the self interactions around realistic flat directions are much more complicated. Each $D-$ term contributes a term of the form $D^aD^a \sim \left( vev \times linear + quadratic \right)^2$ to the potential, where $linear$ and $quadratic$ refer to a sum of terms of that degree in the fluctuations (cf. eq,~(\ref{dalinear}) for a more detailed expression). It is plausible that variances of different fields can grow significantly at preheating, but that the various terms entering in this expression, once summed, will not contribute significantly to the higher terms (more than quadratic) in the potential. Moreover,  the presence of cubic terms in the potential can lead to two possible effects.  Firstly, some MSSM fields may also develop a large vev, as the variances of some other fields are growing at preheating. Secondly, these terms lead to interactions which correspond to the scattering of a quantum against the zero mode and could enhance the decay of the flat direction. These considerations can strongly affect the nonperturbative decay of the flat directions, and they can presumably be studied with the aid of numerical lattice simulations~\cite{latticeeasy}.

\section{Computation of the decay} \label{sec4}

In this Section we return to our original toy model and we show, by explicit computation, that complex flat directions can decay very quickly during their first few rotations. 
We have shown that in more realistic cases, the mass matrices take on a very similar form.
We divide the presentation in three subsections. In the first, we present the general formalism needed for computing the decay. We then perform a numerical computation for a particularly simple background evolution, chosen to single out the specific effect we are considering. In the third subsection, we discuss the computation in a more realistic context. There we will also comment on the case
toy model with two flat directions as we have seen this to be required by simple
counting of degrees of freedom.

\subsection{General formalism} \label{formalism}

To compute the nonperturbative decay of the flat direction $\phi$, one first quantizes the fields $\chi_i$ coupled to it. The quantization occurs in a time-dependent background, due both to the expansion of the Universe - encoded in the scale factor $R$ - and to the evolution of $\phi$ (treated as a homogeneous background field). The time evolution generates quanta of the fields $\chi_i \,$, at the expense of the energy density stored in the coherent motion of $\phi \,$. The standard computation is typically performed with only one field, $\chi$. When more fields are present, there are additional terms, which, as we see below, can be the source of very quick particle production. The general case has been worked out in the second ref. of~\cite{gra2}; here we just present the final results, referring to Section 2 of that paper for a detailed and self-contained derivation.

The system of rescaled fields $R \chi_i$ has the (squared) frequency
\begin{equation}
\Omega_{ij}^2 = R^2 \, {\cal M}_{ij}^2 + \left[ k^2 +  \frac{R''}{R} \left( 6 \xi - 1 \right) \right]
\delta_{ij}
\end{equation}
where $k$ is the comoving momentum, prime denotes a derivative with respect to conformal time $\eta \,$ ($d\eta = dt/R$) and $\xi$ controls the coupling of $\phi$ to the curvature (the value $\xi = 0$ corresponds to minimal coupling, while $\xi = 1/6$ to a conformal coupling). In general, the frequency is not diagonal in field space, due to the nondiagonal mass term. However, we can rotate it through
\begin{equation}
C^T \left( \eta \right) \Omega^2 \left( \eta \right) C \left( \eta \right) = \omega^2 \left( \eta \right) \;\;\;\;\;\; {\rm diagonal}
\end{equation}
We denote by ${\tilde \chi} \equiv C^T \chi$ the fields in the basis in which the frequency matrix is diagonal. We also denote by $\omega_i^2$ the $i-$th entry of the diagonal matrix $\omega^2 \,$. The set of $\omega_i$ represents the frequencies of the (time dependent) physical eigenstates of the system ${\tilde \chi}_i \,$.

Initially, $\Omega^2$ is constant or evolving adiabatically; the fields ${\tilde \chi}_i$ are Fourier decomposed in terms of only positive energy terms, corresponding to a vanishing occupation number.
The time evolution of the frequency also generates negative energy terms; the presence of both positive and negative energy terms signals the production of quanta ${\tilde \chi}_i \,$. In the single field case, the coefficients in front of the positive and negative energy terms, known as Bogolyubov coefficients, are denoted by $\alpha$ and $\beta \,$, respectively. For scalar fields, they satisfy $\vert \alpha \vert^2 - \vert \beta \vert^2 = 1 \,$. The occupation number of the the produced field is given by $\vert \beta \vert^2$ (which is initially vanishing).

In the computation of~\cite{gra2}, the Bogolyubov coefficients are promoted to matrices in field space. They are determined by the initial conditions $\alpha = \mathbbm{1} \;,\; \beta = 0 \,$, and by the evolution equations
\begin{eqnarray}
\alpha ' &=& - i \omega \alpha + \frac{\omega'}{2 \omega} \beta - I \alpha - J \beta \nonumber\\
\beta ' &=& \frac{\omega'}{2 \omega} \alpha + i \omega \beta - J \alpha - I \beta
\label{evolution}
\end{eqnarray}
where we have defined
\begin{equation}
I,J = \frac{1}{2} \left( \sqrt{\omega} \Gamma \frac{1}{\sqrt{\omega}} \pm \frac{1}{\sqrt{\omega}} \Gamma \sqrt{\omega} \right) \;\;\;,\;\;\; \Gamma = C^T \, C'
\end{equation}
(the upper and lower signs refer to $I$ and $J$, respectively). Finally, the occupation number of the $i-$th eigenstate ${\tilde \chi}_i$ can be shown to be
\begin{equation}
n_i \left( \eta \right) = \left( \beta^* \, \beta^T \right)_{ii}
\end{equation}

It is instructive to comment on the evolution equations~(\ref{evolution}). We observe that $\beta$ 
remains zero as long as $\Omega$ is constant. The terms proportional to $\omega' / \omega \,$ are instead the source of particle production in the single field case, where $I=J=0$. They account for the well known gravitational particle production (due to the change of the eigenfrequencies $\omega_i$ with the scale factor $R$), and for preheating (due to the change of the mass matrix ${\cal M}^2 \left( \phi \right) \,$). These terms can be neglected as long as $\omega ' \ll \omega^2$, which was the criterion used in the literature to claim that the nonperturbative decay of flat directions is negligible. However, in the multi-field case, there is an additional source of production, due to the variation of the original frequency matrix $\Omega \,$; this leads to  the terms proportional to $I,J$ in eqs.~(\ref{evolution}). We show below that these terms can lead to the fast nonperturbative decay even if the eigenfrequencies $\omega_i$ are constant or slowly varying.

\subsection{Production in a simple background} \label{toy}

We now apply the formalism given in the previous subsection to a very simple background. We consider the interaction~(\ref{pot2}) between the flat direction $\phi$ and a (complex) field $\chi$, which leads to the mass matrix~(\ref{mass2}). In the next subsection, we discuss the self-consistent and realistic evolution for $\phi$ in a FRW spacetime geometry. Here, instead, we treat them as an external background, chosen so as to single out the production of quanta of ${\tilde \chi}$ from the rotation. Specifically, we work in a Minkowski spacetime, and we take a constant amplitude $\vert \phi \vert \,$; in this way, the $\left\{ {\tilde \chi}_1 ,\, {\tilde \chi}_2 \right\}$ system (real and imaginary components of ${\tilde \chi} \,$, respectively) has constant eigenfrequencies~\footnote{We assume here that the mass for the field $\chi$ arises only from the interaction with the flat direction, that is $m_{{\tilde \chi}_i} = 0$ in eq.~(\ref{mass2}). We discuss the effect of nonvanishing ``bare'' masses at the end of this  subsection.}
\begin{equation}
\omega_1 = \sqrt{k^2 + f^2} \;\;\;,\;\;\; \omega_2 = k
\end{equation}
and particle production is solely due to the varying phase of $\phi$ (encoded in the matrices $I,J$ in eq.~(\ref{evolution})).

The mass matrix ${\cal M}^2$ of the system is given in eq.~(\ref{mass2}), where $\sigma$ denotes the phase of $\phi \,$. We denote by $n_1$ and $n_2$ the occupation numbers of the massive and massless eigenstates, respectively. According to the above discussion, these number densities have an unambiguous meaning only as long as ${\cal M}^2$ is constant (or, at least, adiabatically evolving). For this reason, we assume that $\sigma$ varies significantly only for a finite interval of time, and that it slowly approaches two constant values at early and late times. A simple function with this property is
\begin{equation}
\sigma \left( t \right) = \pi \, N \left[ 1 + \tanh \left( \frac{m_\phi \, t}{\pi \, N} \right) \right]
\label{sigma-toy}
\end{equation}
With this choice, the flat direction performs $N$ complete rotations between $t = - \infty$ and $+ \infty \,$.
Moreover, $\dot{\sigma} = m_\phi$ at its maximum, which is precisely the value that it would assume if it was rotating with constant amplitude in the potential $m_\phi^2 \vert \phi \vert^2  \,$. We interpret $n_i \left( t \rightarrow \infty \right)$ as the occupation number of the $i-th$ eigenstate generated after $N$ rotations of the flat direction. 

To find the occupation numbers, we performed a numerical evaluation of eqs.~(\ref{evolution}). There are two mass scales in the problem given by $\omega_1 \simeq f$ and by $\sigma' \simeq m_\phi \,$. It is convenient to define
\begin{equation}
\mu \equiv \frac{m_\phi}{f} = \frac{m_\phi}{\sqrt{2} g \vert \phi \vert}
\end{equation}
In realistic cases, $\mu$ can be as small as $\sim 10^{-14} \,$. The timescale for particle production is set by the evolution of $\phi \,$, that is $m_\phi^{-1} \,$. However, the intermediate matrices $\alpha ,\, \beta$ evolve on the much quicker timescale $f^{-1} $, dictated by the largest eigenfrequency. Therefore, the computation becomes progressively more time consuming as $\mu$ decreases. We performed evaluations for different values of $\mu \,$, with values of $\mu$ as low as  $10^{-5} \,$. The results obtained exhibit a clear scaling with $\mu \,$, so that the amount of produced quanta can be easily inferred beyond the range we have directly probed.

\begin{figure}[h]
\centerline{
\includegraphics[width=0.7\textwidth,angle=-90]{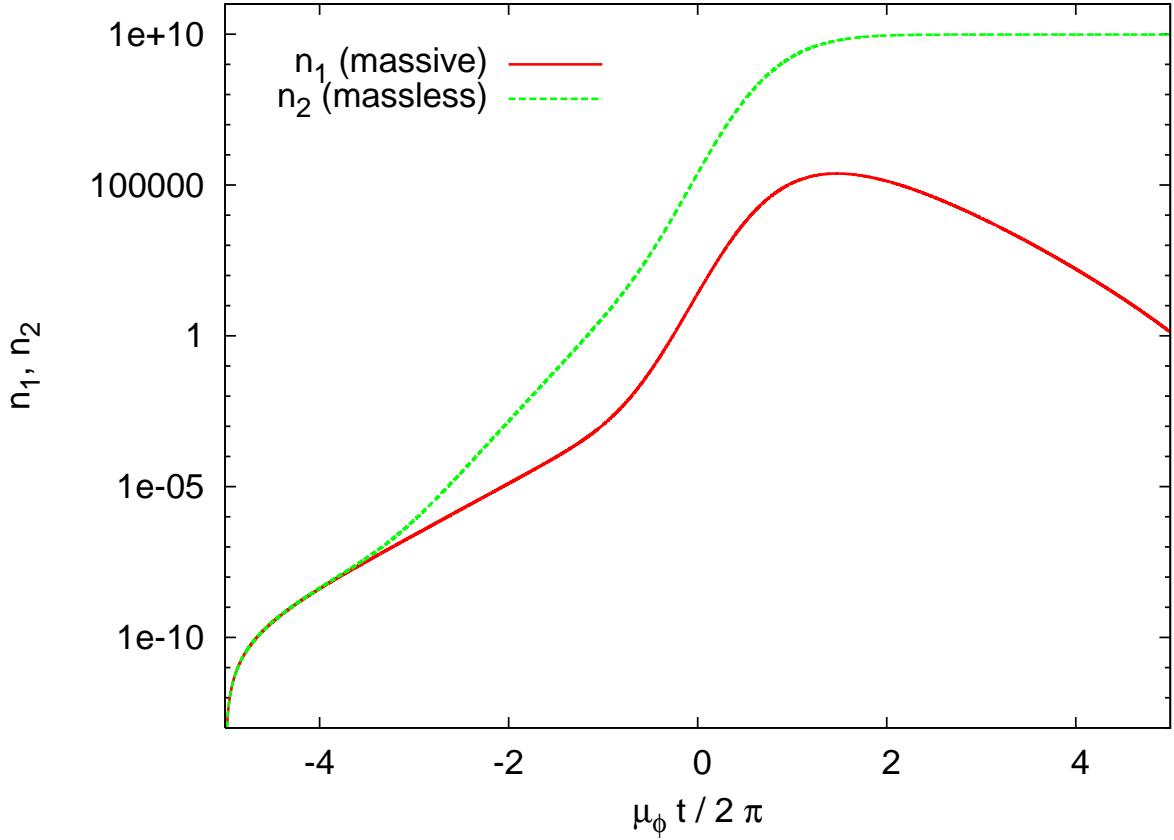}
}
\caption{Growth of the occupation numbers for $2$ rotations, for the specific choice $\mu = 10^{-3}$ and $p = \mu / 10 \,$. The occupation numbers are dimensionless since they represent number densities with respect to both coordinate and momentum space.}
\label{fig1}
\end{figure}

We start the discussion of the numerical results by comparing the growth of the two occupation numbers for a given choice of parameters. We fix $\mu = 10^{-3} \,$ and we consider modes with momentum $k = m_\phi / 10 \,$ (for later convenience, we define the rescaled momentum $p \equiv k / f \,$). The evolution of the two occupation numbers $n_{1,2} \,$ is shown in fig. \ref{fig1}.

In the numerical computations we performed, the argument of the $\tanh$ in~(\ref{sigma-toy}) ranges from $-x_0$ to $+x_0 \,$. The larger $x_0$ is, the more adiabatic the initial and final states are. The results shown in fig. \ref{fig1} have been obtained for $x_0 = 5 \,$. The occupation number for the massless eigenstate grows during the entire evolution, and saturates at a constant value at late times as seen in the figure. In contrast, the occupation number of the massive eigenstate is initially growing, but it eventually decreases as the flat direction comes to a rest. We found that $n_1$ continues to decrease for larger values of $x_0$. Since the mass of the massive eigenstate is much greater than that of the flat direction, we are tempted to argue that the production observed in the figure is an artifact of not having a perfectly adiabatic initial and final state, so that $n_1$ should be vanishing in the limit of $x_0 \rightarrow \infty \,$. This issue, although very interesting per se, does not affect our estimate of the decay time of the flat direction. For any finite value of $x_0 \,$, the numerical results give an upper bound on the amount of massive quanta produced. We verified that even for $x_0 = 5 \,$, the energy density produced in the massless quanta is several orders of magnitude greater then one obtained for the massive state. Therefore, we simply concentrate on the massless quanta, for which  the result is reliable (since $n_2$ has saturated to a constant value).

\begin{figure}[h]
\centerline{
\includegraphics[width=0.7\textwidth,angle=-90]{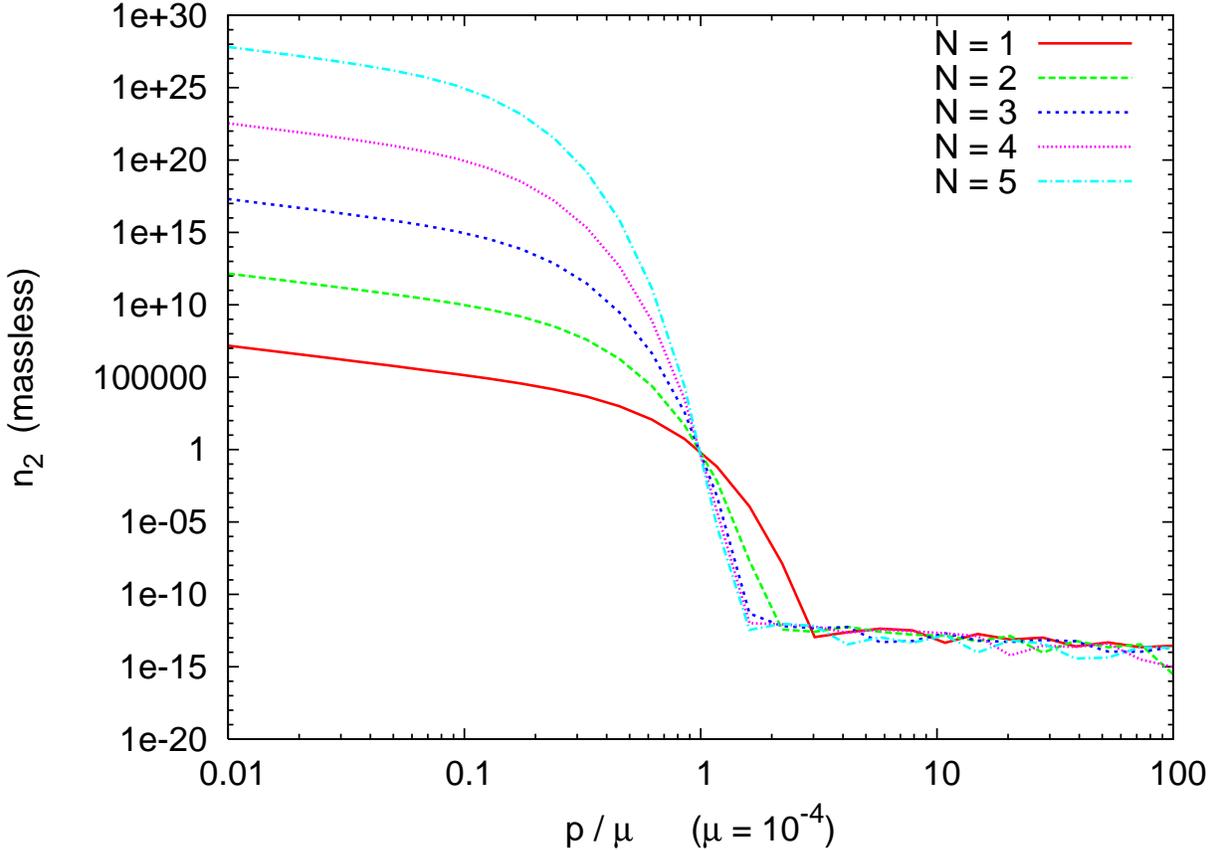}
}
\caption{Spectra of massless quanta produced after $N$ rotations of the flat direction.}
\label{fig2}
\end{figure}

In fig. \ref{fig2},
we show the spectra of massless quanta produced after $N$ rotations of the flat direction. 
The spectra are characterized by an exponentially growing resonance band of momenta lower than the mass of the flat direction, $k < m_\phi \,$. The spectra shown in the figure have been obtained for $\mu = 10^{-3} \,$. We performed an identical study for several values of $\mu \,$, from $10^{-2}$ down to $10^{-5} \,$. We found that the spectra in the resonant band are only a function of $N$ and of the ratio $p / \mu \,$. Therefore, the energy density in the massless quanta, $\rho_{{\tilde \chi}_2} = \int d^3 k \, k \, n_2 \,$, scales as $\mu^4 \,$, and it grows exponentially with $N \,$. The energy density in the flat direction is instead
$\rho_\phi \sim m_\phi^2 \vert \phi \vert^2 \sim \mu^2 f^4 / \left( 2 g^2 \right) \,$. From our numerical results, we find
\begin{equation}
r_\chi \equiv \frac{\rho_{{\tilde \chi}_2}}{\rho_\phi} \simeq 6 \times 10^{-3} g^2 \mu^2 \left( 9 \times 10^4 \right)^N .
\label{ratio}
\end{equation}
This result is reliable only as long as $r_\chi \ll 1 \,$, so that backreaction effects can be neglected. If backreaction is negligible until the end of the process, setting $r_\chi = 1$ in~(\ref{ratio}) provides a good estimate for the decay time of the flat direction. This occurs after
\begin{equation}
N \simeq 6.2 - 0.4 \, {\rm log}_{10} \left( \frac{\mu}{10^{-14}} \right)
\label{ndeca}
\end{equation}
rotations. So, even for realistically small values of $\mu \,$, the nonperturbative decay takes place within a few rotations, on a much shorter timescale than the perturbative one, which would typically last
$\sim 10^{11}$ rotations in a Universe dominated by the radiation from inflaton decays.

As in the case of inflationary preheating, the most important backreaction effect is probably the scattering of the produced quanta against the flat direction. This effect, commonly known as rescattering, destroys the coherent motion of $\phi$, and - consequently - it switches off the resonant production of the quanta of ${\tilde \chi} \,$. So, backreaction decreases the particle production below the estimate~(\ref{ratio}). However, it also contributes to the depletion of the zero mode of $\phi$, so that effectively eq.~(\ref{ndeca}) may still be a good estimate for the decay time of the (coherent) flat direction. We comment more on this issue (and on the following evolution of the system) in the concluding Section.

Finally, we note that a nonvanishing ``bare'' (that is, $\phi-$independent) mass for $\chi$ can be readily included in this computation (we expect it to be present as a soft susy breaking term). A nonvanishing $m_{\chi_1} = m_{\chi_2} = m_\chi$ in eq.~(\ref{mass2}) amounts to replacing $k^2$ with $k^2 + m_\chi^2$ in the computation. So, the parameter $p$ appearing in the above figures is related to the physical momentum $k$ as $k = \sqrt{ \left( p \, f \right)^2 - m_\chi^2} \,$. As a consequence, the spectra shown in fig.~\ref{fig2} should start at $p = m_\chi / f$ rather than at $p = 0 \,$.   The resonance band exists for $\sqrt(p^2+m_\chi^2) < m_\phi$, or $p < \sqrt( m_\phi^2 - m_\chi^2 ) \equiv {\bar p}$.
The resonance band disappears if $m_\chi > m_\phi \,$.
When $m_\chi < m_\phi \,$, the resonance band shrinks to
$0 < p < {\bar p}$   (with ${\bar p} < m_\phi$).
While it is smaller quantitatively, we expect a qualitatively similar
result, unless the masses are nearly degenerate ($m_\chi$
less than but almost equal to $m_\phi$).

\subsection{Production in a cosmological self-consistent background}

We now compute the production of particles in a self-consistent evolving background, characterized by the potential
\begin{equation}
V = \frac{1}{2} m_\phi^2 \vert \phi \vert^2 + \frac{\lambda}{8} \left( \phi^4 + {\rm h. c.} \right) + g^2 \left( \phi \, \chi^* + {\rm h. c.} \right)^2 + \frac{1}{2} m_\chi^2 \vert \chi \vert^2
\end{equation}
The flat direction $\phi \,$ starts to move when the Hubble parameter $H$ decreases below its mass. The evolution is not purely radial towards the origin, but the flat direction acquires some angular motion due to the $\sigma-$ dependent term, proportional to $\lambda \,$. In the Affleck-Dine
mechanism for baryogenesis \cite{ad}, the quartic coupling of interest is found from
a 1-loop baryon number violating interaction with two supersymmetry breaking mass insertions.
As a result, we expect that $\lambda \sim g^4 m_\phi^2/\phi_0^2$ \cite{ad,osch}

In the computation, we rescale $\phi = \phi_0 \, F \,{\rm exp} \left( i \sigma \right) / R \,$, where $\phi_0$ is the initial amplitude of the flat direction (we set $R = F = 1$ as initial conditions). We use conformal time, in inverse units of $f = \sqrt{2} g \phi_0 \,$, and, analogously to $\mu$ and $p$ defined above, we define $\mu_\chi \equiv m_\chi / f \,$. In this notation, the flat direction and the scale factor $R$ evolve according to
\begin{eqnarray}
&& F'' + \left[ R^2 \mu^2 - \sigma'^2 - \frac{R''}{R} \right] F + {\tilde \lambda} F^3 \cos \left( 4 \sigma \right) = 0 \nonumber\\
&& \sigma'' + 2 \frac{F'}{F} \sigma ' - {\tilde \lambda} F^2 \sin \left( 4 \sigma \right) = 0 \nonumber\\
&& \frac{R''}{R} = - \frac{R'^2}{R^2} + 4 \pi  \left\{ f_p^2 \left[ \mu^2 F^2 + {\tilde \lambda} 
F^4 \frac{\cos \left( 4 \sigma \right)}{2 R^2} \right] + \frac{R^2 \rho_\psi}{f^2 M_P^2} \right\}
\label{eom}
\end{eqnarray}
where $\rho_\psi$ is defined in eq. (\ref{rhopsi}) and
dominates the Hubble expansion so long as $\phi_0 \la M_P$. In (\ref{eom}), we have defined $f_p \equiv \phi_0 / M_p \;,\; {\tilde \lambda} \equiv \lambda / 2 g^2 \,$.

We evolved the background equations numerically, starting with $\phi$ at rest due to Hubble friction. It is convenient to define a rescaled Hubble parameter as $H \equiv f \, h \,$, from which we define the parameter, ${\cal F}$:
\begin{equation}
h_I = {\cal F} \mu 
\label{defcalf}
\end{equation}
where $h_I$ is the scaled Hubble parameter after inflation. Using the mass scales described in 
section 1, $m_\psi \sim 10^{-7} M_P$ and $m_\phi \sim 10^{-16} M_P$, we expect ${\cal F} \sim 10^9$.
The larger ${\cal F}$ is, the longer is the initial stage, in which $\phi$ is frozen, and the field $\chi$ is only excited due to the gravitational expansion. Since our focus is on the productions of $\chi$ due to the motion of the flat direction, in the following we present results for ${\cal F} = 10^2 \,$, so that $\phi$ can be consistently started at rest, but the gravitational production of $\chi$ is limited. However, we also discuss below how the final results changes when ${\cal F}$ is increased. For the other parameters, we chose initial conditions leading to sizable rotation. Namely, for any $\mu \,$, we chose
\begin{equation}
{\tilde \lambda} = \mu^2 \;,\; f_p = 0.1 \;,\; \sigma_{\rm in} = 0.2 \;,\; {\cal F} = 10^2
\label{parameters}
\end{equation}

We remind the reader that it is possible that
non-renormalizable operators lead to values of $f$ (and $f_p$)
which are smaller than that assumed for the present calculation.
In geneneral, smaller vevs will lead to a more rapid dissolution
of the condensate as the effective value of $\mu$ will be larger.
Furthermore, as discussed in the introduction, for vevs significantly
below the Planck scale, the flat directions never
dominate the Universe even in the absence of non-perturbative effects.

We can obtain approximate analytical solutions for the background, using the fact that (i) the inflaton oscillations lead to a matter dominated evolution of the scale factor, (ii) that the amplitude of $\phi \,$, decreases as $R^{-3/2}$ during the rotations, and (iii) that the time derivative of the phase is approximately equal to $m_\phi \,$. This leads to
\begin{equation}
R \simeq \left( 3 \pi {\cal F} N \right)^{2/3} \;\;\;,\;\;\; \sigma' \simeq \mu \left( 3 \pi {\cal F} N \right)^{2/3}
\;\;\;,\;\;\; F \simeq \frac{{\cal F}^{2/3}}{\left(3 \pi N \right)^{1/3}}
\end{equation}
where $N$ is the number of rotations of the flat direction  in its complex plane (we will often use $N$ as our ``time variable'', since this leads to a more immediate interpretation of the results). These approximate relations are in good agreement with the numerical results.

The mass matrix for the real and imaginary component of $\chi$ is again of the form~(\ref{mass2}), with $m_{\chi_ 1} = m_{\chi_2} = \mu_\chi \, f\,$. After diagonalization, one finds the two eigenfrequencies
\begin{equation}
\frac{\omega_1}{f} = \sqrt{p^2 + R^2 \mu_\chi^2 +  \frac{R''}{R} \left( 6 \xi - 1 \right) + F^2} \quad\quad,\quad\quad
\frac{\omega_2}{f} = \sqrt{p^2 + R^2 \mu_\chi^2 +  \frac{R''}{R} \left( 6 \xi - 1 \right)} 
\label{eigenfreq}
\end{equation}
while the matrices $I, \, J$ appearing in (\ref{evolution}) acquire the form
\begin{equation}
\frac{I}{f} = \frac{\sigma'}{2}
\, \frac{\omega_1 + \omega_2}{\sqrt{\omega_1 \, \omega_2}} \left( \begin{array}{cc} 0 & - 1\\ 1 & 0 \end{array} \right) \quad\quad,\quad\quad
\frac{J}{f} = \frac{\sigma'}{2} \, \frac{\omega_1 - \omega_2}{\sqrt{\omega_1 \, \omega_2}}
\left( \begin{array}{cc} 0 & - 1\\ -1 & 0 \end{array} \right) 
\label{ij}
\end{equation}

In the case of minimal coupling ($\xi = 0$), the $R''/R \,$ term in $\omega_2$ dominates initially in the range of momenta we are interested in, leading to $\omega_2^2 < 0 \,$. This leads to gravitational production of the second eigenstate. Our focus here is not on the gravitational production, but rather on the creation of quanta of $\chi$ from the motion of the flat direction, which is unrelated to $\xi \,$; moreover, the formalism that we are using assumes that all the eigenfrequencies are real. For these reasons, we assume $\chi$ to be conformally coupled ($\xi = 1/6$). This can be also considered a conservative assumption, since in this way we avoid part of the gravitational creation (there is still gravitational production related to the mass term $R^2 \mu_\chi^2$ in~(\ref{eigenfreq})), so that a conformal coupling results in less particle production relative to the minimally coupled case.

\begin{figure}[h]
\centerline{
\includegraphics[width=0.7\textwidth,angle=-90]{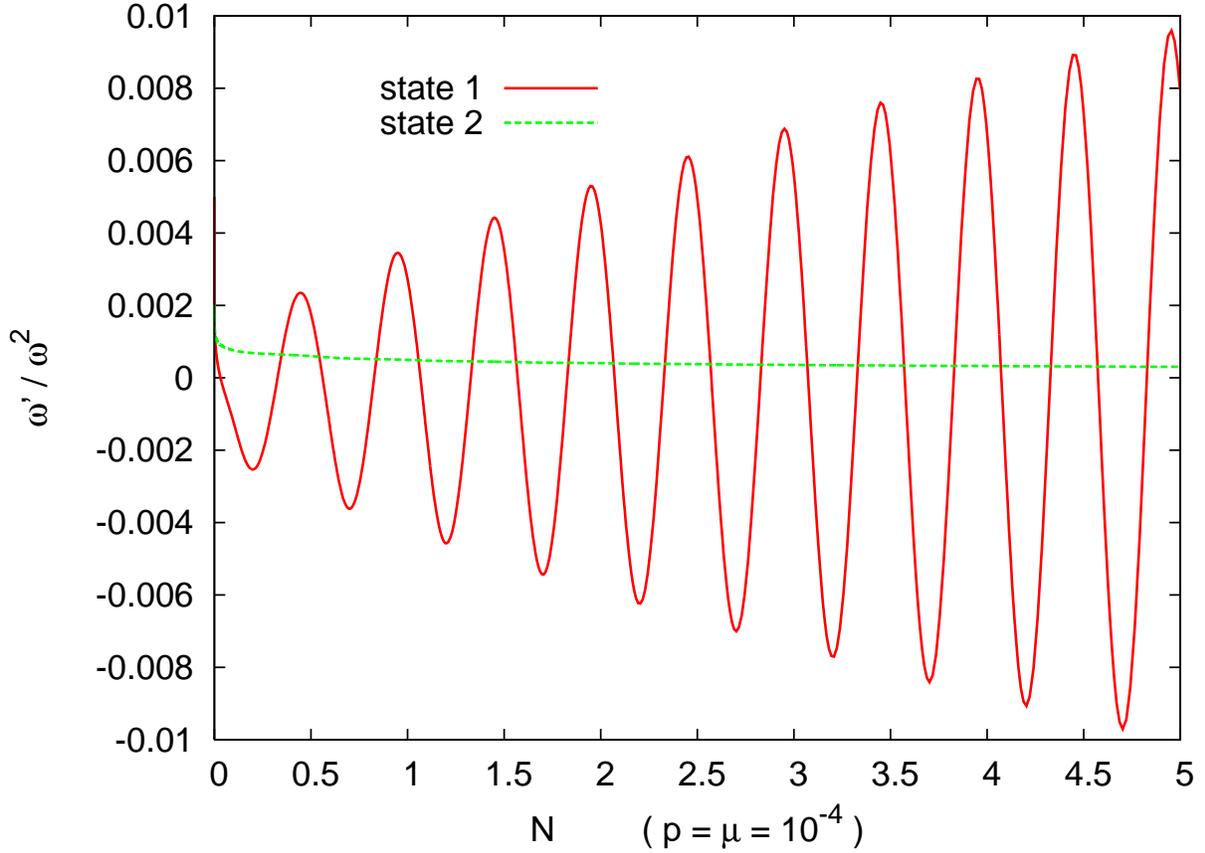}
}
\caption{Evolution of the adiabaticity parameters $ \omega_i' / \omega_i^2 \,$ for the two eigenstates ${\tilde \chi}_1$ and ${\tilde \chi}_2 \,$.}
\label{fig3}
\end{figure}

We start the analysis  by considering a single mode with $p = \mu = 10^{-4} \,$. 
For definiteness, we choose $\mu_\chi = \mu / 2 \,$.
In fig.~\ref{fig3}, we show the evolution of the adiabaticity parameters $\omega_i' / \omega_i^2 \,$.
We note that both $ \omega_i' / \omega_i^2 \ll 1$, corresponding to an adiabatic evolution of the eigenfrequencies. For this reason, one may be tempted to conclude that the nonperturbative production of quanta of $\chi$ is negligible. 

\begin{figure}[h]
\centerline{
\includegraphics[width=0.7\textwidth,angle=-90]{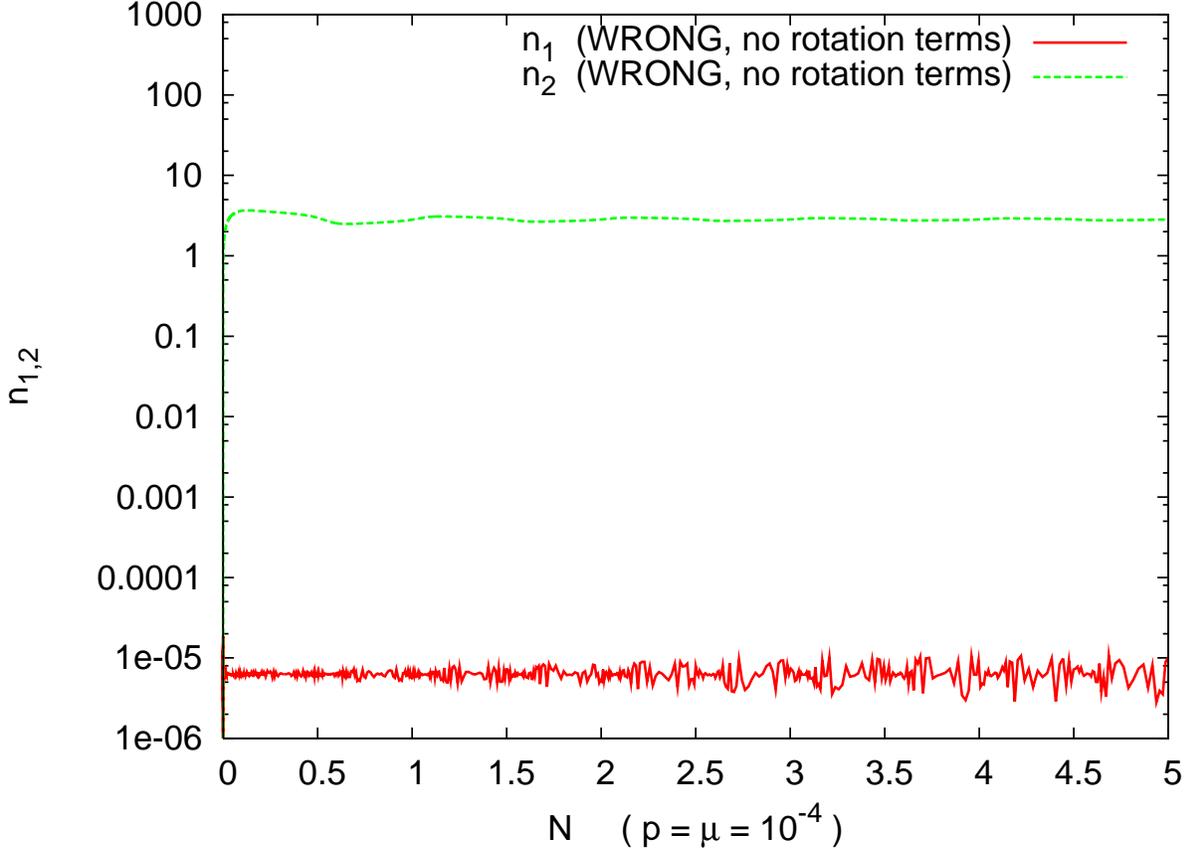}
}
\caption{The incorrect evolution of the occupation numbers ${\tilde \chi}_1$ and ${\tilde \chi}_2 \,$. The matrices $I$ and $J$ are set by hand to zero. Notice there is negligible production, in agreement with the adiabaticity shown in fig. \ref{fig3}.}
\label{fig4}
\end{figure}

\begin{figure}[h]
\centerline{
\includegraphics[width=0.7\textwidth,angle=-90]{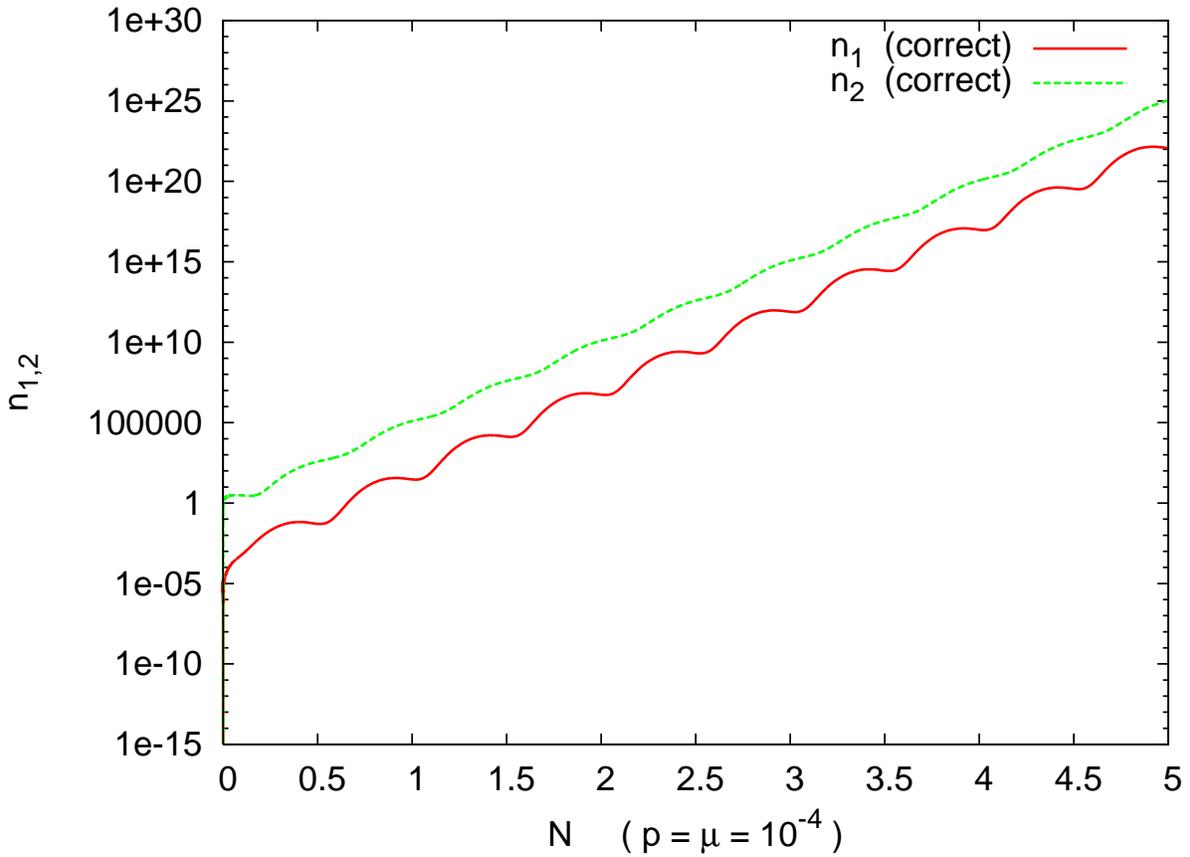}
}
\caption{The correct evolution of the occupation numbers ${\tilde \chi}_1$ and ${\tilde \chi}_2 \,$. Particle production stems from the matrices $I$ and $J$ (now included consistently).}
\label{fig5}
\end{figure}

This conclusion would be correct in the single particle case, or if the mass matrix was diagonal, so that eqs.~(\ref{evolution}) would appear without the matrices $I$ and $J \,$. To verify this, we incorrectly set $I = J = 0$ in these equations, and we show in fig.~\ref{fig4} the evolution of the occupation numbers $n_{i,j}$ of the two eigenstates of the system.
Indeed, we see from the figure that setting $I=J=0$ results in negligible production. In this case, the only source of production would be from the variation of the eigenfrequencies, which, as indicated by fig.~\ref{fig3}, evolve adiabatically. However, when we include the matrices $I$ and $J$ in a manner consistent with eq.~(\ref{ij}), we find significantly more particle production, as shown in fig. \ref{fig5}.

\begin{figure}[h]
\centerline{
\includegraphics[width=0.7\textwidth,angle=-90]{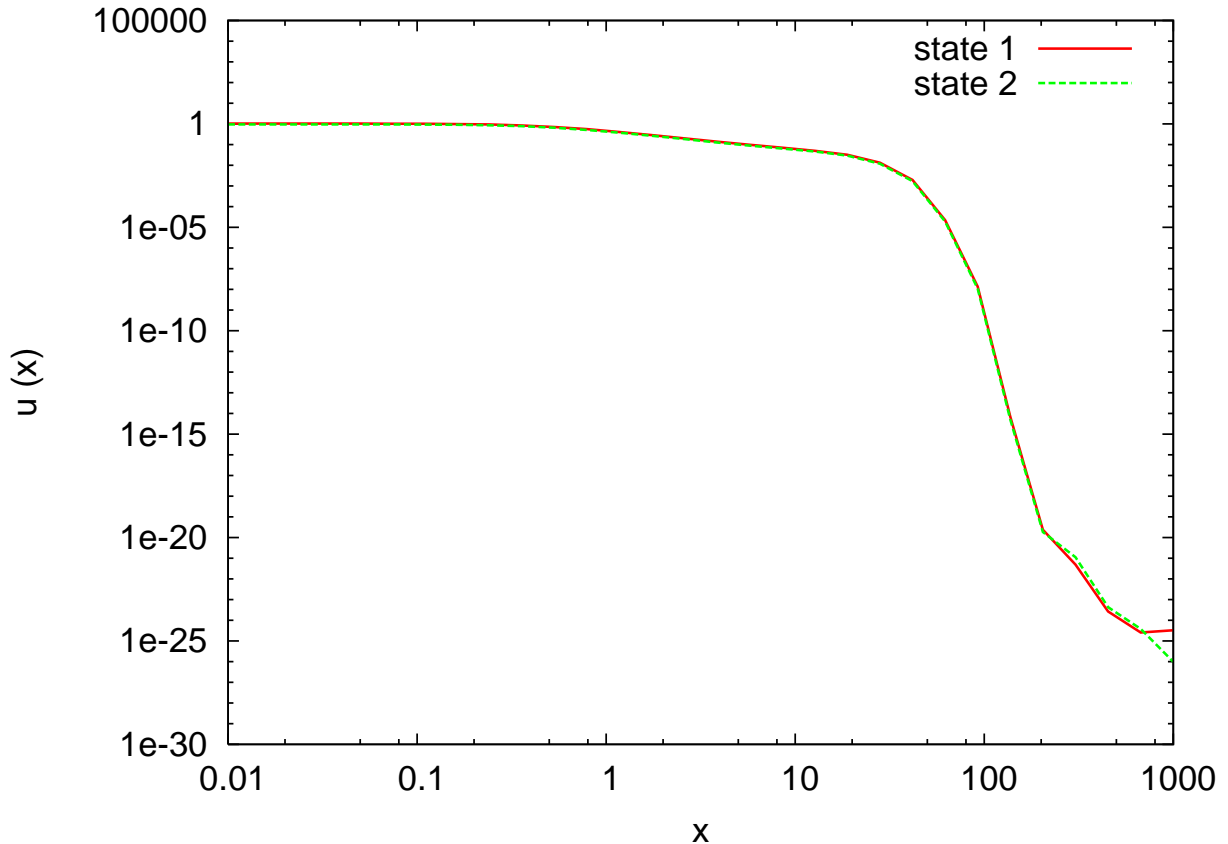}
}
\caption{
Function $u$ entering in eq.~(\ref{scatot}). The two plots have been obtained by numerically evaluating the particle production for $\mu = 10^{-6} \; ,\; {\cal F} = 100 \,$, and $N = 5 \,$ and by scaling the result as in eq.~(\ref{scatot}).}
\label{fig6}
\end{figure}

Having verified the strong effect associated with the rotation of the mass matrix, we can now turn to the computation of the total particle production. We perform the computation for different values of $\mu \,$, 
ranging from~$\mu  = 10^{-4}$ to $\mu = 10^{-7} \,$, studying the production which takes place within the first $N=5$ rotations of $\phi \,$. The spectra of produced quanta show a clear scaling with respect to the parameters of the model:
\begin{eqnarray}
n_1 \left( p ,\, \mu ,\, N ,\, {\cal F} \right) &\simeq& 0.2 \cdot 10.5^{5 N} \cdot \mu \cdot u \left( \frac{p}{\mu} \, \left( \frac{100}{\cal F} \right)^{2/3} \right) \nonumber\\
n_2 \left( p ,\, \mu ,\, N ,\, {\cal F} \right) &\simeq& 0.025 \cdot 10^{5 N} \cdot u \left( \frac{p}{\mu} \, \left( \frac{100}{\cal F} \right)^{2/3} \right) 
\label{scatot}
\end{eqnarray}
where the function $u$ (normalized to $1$ at small $k \,$), is given in fig.~\ref{fig6}.

Among the different scalings summarized in eq.~(\ref{scatot}), the proportionality to $\mu$ is the most accurate. This leads us to conclude that this result can be safely extrapolated to lower (and more realistic) values of $\mu \,$.
With regard to the scaling with the number of rotations $N$, we observe the presence of a resonance band (characterized by exponentially growing occupation numbers) at relatively low momenta. The scaling~(\ref{scatot}) is actually only accurate for the resonance band; however, this is good enough for the computation of the total production, since the quanta produced at higher momenta are completely negligible.

In analogy with the Minkowski case studied in the previous Section, we expect the resonance band to be present for physical momenta smaller than the mass of $\phi \,$, or (in rescaled units), $p / R \la \mu \,$.
Due to the expansion of the Universe, increasingly more comoving momenta enter the resonance band as the scale factor increases. However, the modes which are in the resonant band from the onset of the oscillations of $\phi$ are the ones which experience the most production. 
When the flat direction starts evolving, $R \simeq {\cal F}^{2/3} \,$ (this immediately follows from the definition of ${\cal F} \,$). For ${\cal F} = 100 $, which is the value at which the argument of $u$ is normalized, this amounts to $R \simeq 20 \,$. At the end of the first rotation, we find instead (numerically) $R \simeq 100 \,$. Therefore, for ${\cal F}= 100 \,$, we expect a strong decrease of the occupation numbers in this range of $p / \mu \,$, as fig.~(\ref{fig6}) clearly confirms.

Finally, these considerations also explain the scaling of eqs.~(\ref{scatot}) with respect to ${\cal F} \,$, since ${\cal F}^{2/3}$ is essentially the normalization of the scale factor at the beginning of the oscillations of $\phi \,$, and the results should clearly depend on the physical momenta, rather than on the comoving ones. There is actually another smaller effect of ${\cal F}$ which we have not included in eqs.~(\ref{scatot}). Besides the overall ``shift'' of the spectra that we have just discussed, we also found slightly more production as ${\cal F}$ increases. This is due to gravitational production before $\phi$ starts evolving (the greater ${\cal F}$ is, the greater is the value of the Hubble parameter at the start of the numerical simulation). The result given in (\ref{scatot}) has been obtained for ${\cal F}$ = 100, but - as we mentioned - ${\cal F}$ can be expected to be as large as $10^9$ in realistic cases. Therefore, eqs.~(\ref{scatot}) should be considered as a conservative estimate of the amount of particle production. We also performed numerical computations with ${\cal F} = 10^9 \,$, finding that the total energy density $\rho_\chi$ increases by a factor of $\sim 3 \,$. This however does not change significantly our estimate for the moment at which the flat direction decays (see eqs.~(\ref{ratfc}) and  (\ref{ndeca2}) below).

After this discussion, we can now compute the energy density of quanta of $\chi$ produced, and its ratio with the energy density stored in the coherent oscillations of $\phi \,$. We estimate the decay time as the moment at which this ratio becomes one (with all the cautionary remarks already made at the end of the previous subsection).

In the resonance band, and in the time interval we are interested in, the eigenfrequencies~(\ref{eigenfreq}) can be estimated as
\begin{eqnarray}
\frac{\omega_1}{f} &\simeq& F \sim \frac{{\cal F}}{\left( 3 \pi {\cal F} N \right)^{1/3}} \nonumber\\
\frac{\omega_2}{f} &\simeq& R \, \mu_\chi \sim \frac{\mu}{2} \, \left( 3 \pi {\cal F} N \right)^{2/3} 
\end{eqnarray}

This, together with eq.~(\ref{scatot}) for the occupation numbers, leads to
\begin{eqnarray}
\rho_\chi &\simeq& \frac{1}{R^4} \int d^3 k \left( \omega_1 \, n_1 + \omega_2 \, n_2 \right) \\
&\simeq& \frac{4 \pi f^4}{R^4} \mu {\cal F}^{2/3} \left[ \frac{1}{\left( 3 \pi N \right)^{1/3}} 0.2 \cdot 10.5^{5N} + \frac{\left( 3 \pi N \right)^{2/3}}{2} 0.025 \cdot 10^{5N} \right] \int d p \, p^2 \, u \nonumber
\end{eqnarray}
the energy density of the $2$nd state dominates in the range of $N$ we are interested in; the integral over momenta is numerically found to be $\int d p \, p^2 u \simeq 0.03 \, \mu^3 {\cal F}^2 \,$. So,
\begin{equation}
\rho_\chi \simeq 5 \cdot 10^{-3} \frac{f^4 \mu^4 {\cal F}^{8/3}}{R^4} \left( 3 \pi N \right)^{2/3} 10^{5 N}
\simeq 5 \cdot 10^{-3} \frac{f^4 \mu^4 {\cal F}^2}{R^3} 10^{5 N}
\end{equation}

For the coherent motion of the flat direction, we have instead 
\begin{equation}
\rho_\phi \simeq  m^2 \vert \phi \vert^2 = f^2 \mu^2 \phi_0^2 \frac{F^2}{R^2}
\end{equation}
Using the above estimates for $F$ and $R \,$, 
\begin{equation}
\rho_\phi \simeq \frac{f^4 \mu^2}{2 g^2 R^3} {\cal F}^{2} 
\end{equation}
giving the ratio
\begin{equation}
\frac{\rho_\chi}{\rho_\phi} \simeq 10^{-2} g^2 \mu^2 10^{5 N}
\label{ratfc}
\end{equation}
Notice that ${\cal F}$ cancels in the ratio, since its role is simply related to the value of the scale factor at the beginning of the rotations. For $g^2 \sim 0.4 \,$, the ratio equates to $1$ for
\begin{equation}
N_{\rm decay} \simeq 6.1 - 0.4 \, {\rm log }_{10} \left( \mu / 10^{-14} \right)
\label{ndeca2}
\end{equation}
From this we conclude that the flat direction decays long before the inflaton field.

Finally, we return to the our enhanced toy model with two flat directions.
Here, even in the gauged case, we expect mixing between four degrees of freedom
(the other four are decoupled and represent the two flat directions).  We recall however,
that only production of the two light degrees of freedom which are orthogonal to the
Goldstone mode can be responsible for preheating. 
Indeed, when we generalize the analysis detailed above, 
and when the two flat directions are characteriized by two independent
backgrounds, $\phi = |\phi|e^{i \sigma}$ and $\phi^\prime = |\phi^\prime| e^{i \theta}$,
we find that the occupation numbers, $n_1, n_2, n_3$ and $n_4$
are all produced as in Fig.~\ref{fig5}.  Furthermore, we have checked explicitely that
when the two flat directions are degenerate, and $\phi = \phi^\prime$, $n_3 = n_4 = 0$
and we return to the case where only the Higgs and Goldstone modes 
are involved.  Thus we require two independent flat directions to be excited in order
for preheating to occur.  As we saw in section 3, this feature is rather common in the MSSM.

\section{Summary and Conclusions} \label{sec5}

The scalar potential of the MSSM is characterized by several flat directions. This flatness may be lifted 
by supersymmetry breaking terms, nonrenormalizable operators, and/or supergravity corrections. The latter are of particular relevance in the early Universe. For instance, if the MSSM fields have a minimal Kahler potential, then supergravity terms provide a mass (of order H) for the flat directions which prevents them from being excited during inflation \cite{drt}. However, if the Kahler potential exhibits a
Heisenberg symmetry \cite{BG}, such mass terms do not arise at the tree level and the flat directions can indeed develop a large vev during inflation \cite{gmo}. No-scale supergravity \cite{ns}
is one such example. The development of a large vev may have interesting consequences for the subsequent evolution of the Universe.

Among the consequences of a large vev along a supersymmetric flat direction, is the possible generation of a baryon asymmetry \cite{ad}. Furthermore,  the energy density stored along the flat direction may come to dominate over the inflaton decay products if the flat direction is long-lived. It is tempting to conclude that this is the case if one simply considers the perturbative (single quanta) decay of the flat direction. Indeed, the flat direction, due to its large vev, provides a large mass to the fields to which it is coupled, so that the perturbative decay proceeds by loop(s), with very heavy fields in the propagator. This typically results in a small decay rate.

However, this consideration does not rule out the possibility that the flat directions decay
nonperturbatively. It has been shown \cite{kls}, that in many models of inflation, the inflaton decays nonperturbatively through a collective effect called preheating. In this paper, we have investigated the possibility of preheating during the evolution of fields associated with supersymmetric flat directions. In particular, it is important to determine whether preheating can lead to a significantly quicker decay than the perturbative estimates may suggest. Prior to our investigation, this question was addressed only in a very limited number of papers, which concluded that preheating is negligible in this case. Their claim, however, was based on toy models, where the flat direction $\phi$ is coupled to a single scalar $\chi$ through a quartic $\vert \phi \vert^2 \vert \chi \vert^2$ interaction. The time varying amplitude of $\phi$ results in a time varying mass for the real and imaginary components of $\chi \,$. However, in the specific case in which $\phi$ is a flat direction (rather than the inflaton) the variation of this mass is always too slow to lead to significant nonperturbative particle production.

We have shown, however,  that the couplings of the flat direction encoded in the MSSM potential are more complicated than the simple $\vert \phi \vert^2 \vert \chi \vert^2$ interaction. In particular, the part of the potential from the $D-$terms results in couplings of the form $\phi^2 \chi_i \chi_j$, leading a non-diagonal mass matrix for the fields $\chi_i$. Even if the eigenmasses are slowly varying, the mass matrix ${\cal M}^2$ itself varies with time. The dominant time dependence is due to the phase $\sigma$ of the flat direction, which is rotating quasi periodically (with a small decrease of the amplitude) around the minimum of its potential. The quasi periodic variation of ${\cal M}^2$ leads to a very strong parametric resonance effect. We found that when at least two flat directions are excited, the flat directions typically loses most of their energy density within its first $5-10$ rotations due to these couplings. 
From this point on, backreaction effects of the $\chi_i$ fields can be expected to play a very relevant role in the evolution of the system. Their complete investigation requires the use of numerical simulations on the lattice \cite{latticeeasy}, which are beyond the aim of the present work\footnote{
Numerical simulations are also required to determine if non-linear interactions 
affect the early stages of preheating.}.
However, due to the strong interactions between the MSSM quanta, these interactions are likely to lead to a fast depletion of the (remaining) vev of the flat direction. As such, it is very unlikely that the scalar field oscillations of the flat direction can ever come to dominate over the inflaton field and its decays product, particularly in the case in which the inflaton itself decays only gravitationally.

Although the final stage of the decay remains to be explored, we have thus showed that flat directions can potentially 
have a very fast decay channel which has been so far overlooked in the literature.  
There are several 
important consequences of this rapid decay.  If flat directions never
dominate the energy density of the Universe, entropy production and reheating are
determined by the dynamics of the inflaton. 
The baryon asymmetry will be measured with respect to 
the entropy density produced by inflaton decays and will typically be large
requiring either further dilution of the baryon asymmetry by later entropy production (e.g.
due to moduli decays) or regulating the initial vev along the flat direction with
some non-renormalizable operator \cite{cgmo}. Finally, in the absence of a large 
vev and hence large particle masses, inflaton decay and the subsequent thermalization
also proceeds rather quickly \cite{ds},
leading to a relatively large reheat temperature after inflation which could
have dire consequences on the gravitino problem.  

\vspace{1cm}
\noindent{ {\bf Note added} } : After the present work was completed, a new paper which discusses the decay of flat directions \cite{am2} came to our attention. The first part of \cite{am2} is devoted to the case of a single flat direction. The analysis, and the conclusions, coincide with those already given in the present work.
We comment on two of the remarks made in the second part (devoted to the case of two or more flat directions), although neither of them has been supported by a concrete computation. The first remark claims that flat directions may have hierarchical vevs (a small ratio $r$ between the two vevs),
so that the system may be effectively reduced to the single flat direction case.
In general, the amount of produced particles will be suppressed by the ratio $r$. However, for any finite $r$, the number densities of the non-goldstone massless fields still grow exponentially with the number $N$ of rotations of the flat directions. For the toy model with two flat directions, we have verified numerically that most of the energy density in the flat directions decays after $N \sim 10$ when $r = 10^{-3}$ (as opposed to $N \sim 5$ when $r=1$).
The second remark is that nonrenormalizable terms in the potential, which depend also on the phase of the flat directions,
can in some cases result in chaotic motion for the phases, potentially suppressing the parametric amplification which takes place when the
phases move with nearly constant velocity.
First, nonrenormalizable interactions are very model dependent, and do not necessarily lead to chaotic motion for the phases.
Second, even if they do, nonrenormalizable terms are relevant only for the first few oscillations of the flat directions. As soon as the flat directions evolve, their vev is decreased by the expansion of the universe, entering into a region where their potential can be safely approximated by the quadratic (phase-independent) part. When this happens, the phases will evolve with a nearly constant velocity, as the cases considered here.

\vskip 0.5in
\vbox{
\noindent{ {\bf Acknowledgments} } \\
\noindent
M.P. thanks Lorenzo Sorbo for collaboration in obtaining the formalism used in this paper and in its study. We also thank Joel Giedt, Lev Kofman,  Erich Poppitz, and Misha Shifman for very useful discussions. We would particularly like to thank Rouzbeh Allahverdi for very pertinent comments 
on the original version of this manuscript.
This work was partially supported by DOE grant DE--FG02--94ER--40823. M.P. also 
acknowledges a grant from the Office of the Dean of the Graduate School of the University of Minnesota.  
}

\end{document}